\documentclass[draft]{article}
\usepackage{amsmath,amsfonts,latexsym, amssymb}

\usepackage{color}

\newcommand{\e}{\varepsilon}

\newcommand{\pt}{\partial}
\newcommand{\rd}{{\rm d}}
\newcommand{\bR}{{\mathbb R}}
\newcommand{\bbZ}{{\mathbb Z}}

\newcommand{\ba}{{\bf{a}}}

\newcommand{\bx}{{\bf{x}}}

\newcommand{\bu}{{\bf{u}}}
\newcommand{\bv}{{\bf{v}}}

\newcommand{\wh}{\widehat}

\newcommand{\al}{\alpha}

\newcommand{\be}{\begin{equation}}
\newcommand{\ee}{\end{equation}}

\newcommand{\om}{{\omega}}

\newcommand{\cL}{{\cal L}}

\newcommand{\cN}{{\cal N}}

\newcommand{\cH}{{\cal H}}

\newcommand{\im}{{\text{Im} }}

\newcommand{\E}{{\mathbb E }}
\newcommand{\R}{{\mathbb R }}

\renewcommand{\P}{{\mathbb P}}

\newcommand{\wt}{\widetilde}

\newcommand{\tr}{\mbox{Tr}\, }

\newtheorem{theorem}{Theorem}

\newtheorem{lemma}[theorem]{Lemma}

\newcommand{\qed}{\hfill\fbox{}\par\vspace{0.3mm}}

\numberwithin{equation}{section}
\numberwithin{theorem}{section}
\numberwithin{definition}{section}
\title{Universality of Wigner Random Matrices}

\author{L\'aszl\'o Erd\H os${}^1$\thanks{Partially supported
by SFB-TR 12 Grant of the German Research Council}
  \\ \\
Institute of Mathematics, University of Munich, \\
Theresienstr. 39, D-80333 Munich, Germany}
\begin{document}
\date{Sep 15, 2009}

\maketitle

\begin{abstract}

We consider  $N\times N$ symmetric or hermitian random matrices 
with independent, identically distributed
 entries where the probability distribution  for each matrix element is given by 
a measure $\nu$ with a subexponential decay. 
We prove that the local eigenvalue statistics in the bulk of the spectrum for 
these  matrices coincide with those of the Gaussian Orthogonal Ensemble (GOE)
and the Gaussian Unitary Ensemble (GUE), respectively, in the limit $N\to \infty$.
Our approach is based on the study 
of the Dyson Brownian motion via a related new dynamics, the local relaxation flow.
We also  show that the Wigner semicircle law holds locally
on the smallest possible scales and we prove that eigenvectors
are fully delocalized and eigenvalues repel each other on arbitrarily small
scales.
\end{abstract}


\medskip


\section{Introduction} 

A central question concerning random matrices is the universality conjecture 
which states that  local statistics  of eigenvalues of large 
$N\times N$ square matrices $H$ 
are determined by  the symmetry type of the ensembles 
but are otherwise  independent of the details of  the distributions.

There are two types of universalities: the edge universality and the
 bulk universality concerning the interior of the spectrum.  The edge 
universality is commonly approached via the fairly robust 
moment method \cite{SS, Sosh}; very recently an alternative
proof was given \cite{TV2}.
 The bulk universality is a subtler 
problem. In the hermitian case, it states that the local $k$-point correlation functions
of the eigenvalues, after appropriate rescaling, are given
by the determinant of the  {\it sine kernel}
\be
  \det \big( K(x_\ell - x_j)\big)_{\ell,j=1}^k, \qquad K(x) = \frac{\sin \pi x}{\pi x},
\label{sine}
\ee
independently of the distribution of the entries.
Similar statement holds for the symmetric matrices
but the explicit formulae are somewhat more complicated.

For ensembles that remain invariant under the
transformations $H\to U^*HU$ for any unitary matrix $U$, the joint
probability density function of all the $N$ eigenvalues can be explicitly
computed.  These ensembles are typically given by the
probability density
$$
    P(H)\rd H \sim \exp(-N\tr V(H)) \rd H
$$
where $V$ is a real function with sufficient growth at 
infinity and $\rd H$ is the flat measure.
The eigenvalues are strongly correlated and they
 are distributed according to a Gibbs measure
with a long range logarithmic interaction potential. 
The joint probability density of the eigenvalues of $H$
can be computed explicitly:
\be
    f(\lambda_1, \lambda_2, \ldots \lambda_N) 
  = \mbox{const.} \prod_{i<j} (\lambda_i-\lambda_j)^\beta \prod_{j=1}^N
   e^{- N\sum_{j=1}^N V(\lambda_j)},
\label{expli}
\ee
where $\beta=1$ for symmetric and $\beta=2$ for hermitian ensembles.
The local
statistics can be obtained via a detailed analysis of
orthogonal polynomials on the real line with respect to the 
weight function $\exp(-V(x))$. Quadratic $V$ corresponds to
the Gaussian ensembles.
 This approach was originally applied \cite{M}
for ensembles that lead to  classical orthogonal polynomials
(e.g. GUE leads to Hermite polynomials). Later  a
general method using orthogonal polynomials has been developed to tackle a very general
class of unitary ensembles 
(see, e.g. \cite{BI, DKMVZ1, DKMVZ2, M, PS} and references therein).

Many natural matrix ensembles are typically not unitarily invariant; the most
prominent example is the Wigner matrices. These are
symmetric or hermitian  matrices whose  entries above the diagonal  
are independent,  identically distributed random variables. 
The only  unitarily invariant Wigner ensembles 
are the Gaussian ensembles. For general Wigner matrices,
no explicit formula is available for the joint eigenvalue
distribution. Thus the basic algebraic connection between
eigenvalue ensembles and orthogonal polynomials is missing
and completely new methods needed to be developed.

The bulk universality for  {\it hermitian} Wigner ensembles
has been established  jointly with S.  P\'ech\'e,  J. Ramirez, 
  B. Schlein and H.T, Yau,
and independently  by Tao-Vu
\cite{ERSY2, TV, ERSTVY}. These works rely on the Wigner
matrices with Gaussian divisible distribution, i.e. 
ensembles  of the form 
\be
\wh H+ s V,
\label{HaV}
\ee
where $\wh H$  is a Wigner matrix,
$V$ is an independent standard GUE matrix and $s$ is a positive constant. 
Johansson  \cite{J} (see also \cite{BP}) proved bulk universality for
the eigenvalues of such matrices using an
{\it explicit} formula by Br\'ezin-Hikami  \cite{BH, J}
on the correlation functions. 
Unfortunately, the similar formula for symmetric matrices
 is not very explicit and the technique of \cite{ERSY2, J}
cannot be extended to prove universality for symmetric
Wigner matrices.

A key observation of Dyson is that if the  parameter $s$ in the matrix $\wh H+ s V$
is varied 
and $s^2$ is interpreted as time, then 
 the evolution of the eigenvalues is given by a coupled system of
stochastic differential equations, commonly called
  the  Dyson Brownian motion (DBM) \cite{Dy}. 
If we replace the Brownian motions by the 
Ornstein-Uhlenbeck  processes to keep the variance constant,
then the resulting dynamics on the eigenvalues,
 which we still call
 DBM,  has the GUE eigenvalue  distribution as the invariant measure. 
Thus the result of Johansson can be interpreted as stating that the 
local statistics of GUE is
 reached via DBM for time of order one. In fact, by analyzing the dynamics 
 of DBM with ideas from the hydrodynamical limit, 
 we have extended Johansson's result  to $s^2\gg N^{-3/4}$  \cite{ERSY}.
 The  key 
observation of \cite{ERSY}   is that the local statistics of 
eigenvalues  depend exclusively on the approach to local equilibrium
which in general is faster than reaching global equilibrium.
Unfortunately, the identification of  local equilibria  still uses 
explicit representations of correlation functions by  orthogonal polynomials
(following e.g. \cite{PS}),   and the extension to other ensembles is not a simple task. 

Therefore, the universality for symmetric random matrices  remained
open and the only partial result is  given  by Tao-Vu (Theorem 23 in \cite{TV})  for
Wigner matrices with  the first four moments of the matrix elements 
matching  those of GOE.

In \cite{ESY4},  together with B. Schlein and H.T. Yau,
we have introduced a general approach 
based on a new stochastic flow,  the local relaxation flow, which locally behaves like 
DBM, but has a faster decay to equilibrium.  This approach
completely circumvents explicit formulae. It  
is thus applicable to prove the universality  for a very broad
class of matrices that includes hermitian, symmetric and symplectic
Wigner matrices and in principle it works also for Wishart matrices
and general $\beta$-ensembles. The heart of the
proof is a convex analysis and the model specific information
involve only estimates on the
accuracy of the local density of states.
For simplicity of the presentation, we will focus on
the hermitian and symmetric cases, the necessary modifications for
the other cases are technical.

We present results 
only about the convergence of the local correlation functions; 
this implies, among others, that the distribution of the gap (difference between neighboring
eigenvalues) is universal as well (Wigner surmise). In particular, short gaps are suppressed,
i.e. the eigenvalues tend to repel  each other. This feature is characteristic
to the strongly correlated point process of  eigenvalues of random matrices
in contrast to the Poisson process of independent points.

\bigskip

Universality of local eigenvalue statistics is believed to hold
for a much broader class of matrix ensembles than we have introduced.
Wigner has originally invented random matrices to mimic the eigenvalues
of the then unknown Hamiltonian of heavy nuclei; lacking any
information, he assumed that the matrix elements are i.i.d.
random variables subject to the hermitian condition. Conceivably, the matrix
elements need not be fully independent or identically distributed for universality.
There is little known  about matrices with 
correlated entries, apart from the unitary invariant ensembles that represent a very specific 
correlation. In case of a certain class of Wigner matrices with weakly correlated
entries, the semicircle law and its Gaussian fluctuation
have been proven \cite{SSh1, SSh2}.  

Much more studied are
various classes of random matrices with independent but not identically distributed entries.
The most prominent example is the Anderson model \cite{A}, i.e. a Schr\"odinger operator
on a regular square lattice with a random potential. Restricted to a finite box,
it can be represented by a matrix whose diagonal elements are i.i.d. random variables;
the deterministic off-diagonal elements are  given by the Laplacian.
In space dimensions three or higher and for weak randomness,
the Anderson model is conjectured to exhibit metal-insulator transition.
Near the spectral edges, the eigenfunctions are localized \cite{FS, AM}
and the local eigenvalue statistics is Poissonian \cite{Mi}; in particular there is 
no level repulsion.  It is conjectured, but not yet proven, that 
in the middle of the spectrum the eigenfunctions
are  extended (some results on the quantum diffusion and delocalization
of eigenfunctions are available in a
certain scaling limit \cite{ESY, Ch}).
 Furthermore, in the delocalization regime the local
eigenvalue statistics are expected to be given by GUE or GOE statistics,
depending whether the time reversal symmetry is broken by magnetic field or not.
Based upon this conjecture, local eigenvalue statistics is
used to compute the phase diagram numerically.
It is very remarkable that the random Schr\"odinger operator,
represented by a very sparse random matrix, exhibits
the same universality class as the full Wigner matrix, at least in
a certain energy range. 

An intermediate class of
ensembles between these two extremes is the family of random band matrices.
These are hermitian or symmetric random matrices $H$ with 
independent but not identically distributed entries. The variance
of $H_{ij}$ depends only on $|i-j|$ and it becomes negligible
if $|i-j|$ exceeds a given parameter $W$, the band-width; for example,
$\E |H_{ij}|^2 \sim \exp(-|i-j|/W)$. It is conjectured
that for narrow bands, $W\ll \sqrt{N}$, the local eigenvalue
statistics is Poisson, while for broad bands, $W\gg \sqrt{N}$
it is given by GUE or GOE, depending on the symmetry class.
(Localization properties of $H$ for $W\ll N^{1/8}$
has been recently shown \cite{Sch} but not local statistics.)
To mimic the three dimensional Anderson model, the rows
and columns of $H$ may be labelled by a finite domain of 
the three dimensional lattice, $i,j \in \Lambda\subset\bbZ^3$.
The only rigorous result for this three dimensional band matrix concerns
the density of states by establishing that Wigner semicircle law holds
as $W\to\infty$ \cite{DPS}.

Finally, we mention that universality of local eigenvalue statistics is often
investigated by supersymmetric techniques in the physics literature.
 These methods are extremely powerful
to extract the results by saddle point computations, 
but the analysis justifying the saddle point
approximation usually lacks mathematical rigor. It is a challenge
to the mathematical physics community to put the supersymmetric
method on a solid mathematical basis; so far only the density
of states has been investigated 
rigorously by using this technique  \cite{DPS}.

\section{Local semicircle law, delocalization and level repulsion}\label{sec:sc}

Each approach that proves bulk universality for general Wigner matrices
requires first to analyze the local density of eigenvalues.
The Wigner semicircle law \cite{W} (and its analogue for
Wishart matrices, the Marchenko-Pastur law \cite{MP}) has traditionally been
among the first results established on random matrices.
Typically, however, the empirical density  is shown to
converge weakly on macroscopic scales, i.e.
on intervals that contain $O(N)$ eigenvalues. 
Based upon our results \cite{ESY1, ESY2, ESY3}, here we show that
the semicircle law holds on much smaller scales as well.

To fix the notation, we assume that in the symmetric case
the matrix elements  of $H$ are given by
\be
   h_{\ell k} = N^{-1/2}    x_{\ell k}, 
\label{scaling}
\ee
where $x_{\ell k}$ for $\ell<k$ are independent,
identically distributed random variables
with the distribution $\nu$ that has
zero expectation and  variance $1$.  The diagonal
elements  $x_{\ell \ell}$ are also i.i.d.  with distribution $\wt \nu$
that has zero expectation and  variance two. 
In the hermitian case we assume that
\be
   h_{\ell k} = N^{-1/2} (x_{\ell k}+iy_{\ell k})
\label{herm}
\ee
where $x_{\ell k}$ and $y_{\ell k}$ are real i.i.d.  
random variables with zero expectation and variance $\frac{1}{2}$.
The diagonal elements are also centered and have variance one.
The eigenvalues of $H$ will be denoted by $\lambda_1\le \lambda_2 \le 
\ldots \le \lambda_N$. The Gaussian ensembles (GUE and GOE) are
special Wigner ensembles with Gaussian single-site distribution.

We will often need to assume that the distributions
$\nu$ and $\wt\nu$ have Gaussian decay, i.e. there exists
$\delta_0>0$ such that
\be
      \int_\R \exp \big[ \delta_0 x^2\big] \rd\nu(x) < \infty,
\qquad \int_\R \exp \big[ \delta_0 x^2\big] \rd\wt\nu(x) < \infty.
\label{gauss}
\ee
In several statements we can relax this condition
to assuming only subexponential decay, i.e. that
there exists $\delta_0>0$ and $\gamma>0$ such that
\be
      \int_\R\exp \big[ \delta_0 |x|^\gamma\big] \rd\nu(x) < \infty,
\qquad \int_\R\exp \big[ \delta_0 |x|^\gamma\big] \rd\wt\nu(x) < \infty.
\label{subexp}
\ee

The matrix elements have thus variance of order $1/N$.
This normalization guarantees
that the spectrum remains bounded as $N\to \infty$,
in fact the spectrum converges to $[-2,2]$ almost surely.
Therefore the typical spacing between neighboring eigenvalues is of
order $1/N$.

For any $I\subset \R$ let $\cN_I$ denote the number
of eigenvalues in $I$. Wigner's  theorem \cite{W} states
that for any fixed interval $I$
$$
    \frac{\cN_I}{N} \to \int_I \varrho_{sc}(x) \rd x
$$
almost surely as $N\to\infty$, where
$$
   \varrho_{sc}(x) : =\frac{1}{2\pi}\sqrt{(4-x^2)_+}
$$
is the density of the semicircle law. This result can be
interpreted as a law of large numbers for the 
empirical eigenvalue density on macroscopic scales, i.e.
for intervals that contain $O(N)$ eigenvalues. 
The following result shows that the semicircle law holds
on intervals $I$ of length $|I|=\eta \ge K/N$
for sufficienly large $K$.

\begin{theorem}\label{thm:locsc} \cite[Theorem 3.1]{ESY3}
Suppose that \eqref{gauss} holds. Let $\kappa>0$ and
fix an energy $E\in[-2+\kappa, 2-\kappa]$. Consider
the interval $I= \big[ E -\frac{\eta}{2}, E+\frac{\eta}{2}\big]$
of length $\eta$ about $E$. Then
there exist  positive constants $C, c$, depending
only on $\kappa$, and a universal constant $c_1$ such
that for any $\delta \le c_1\kappa$ there is $K=K_\delta$
such that
\be
 \P\Big\{ \Big| \frac{\cN_I}{N\eta} - \varrho_{sc}(E)\Big|\ge \delta\Big\}
  \le C e^{-c\delta^2 \sqrt{N\eta}}
\label{P}
\ee
holds for all $\eta$ satisfying $K/N\le \eta\le 1/K$.
\end{theorem}

In particular, this result shows that $\cN_I/N\eta$ converges
to $\varrho_{sc}(E)$ in probability as long as $\eta=\eta(N)$ is
such that $\eta(N)\to 0$ and $N\eta(N)\to \infty$. The Gaussian
decay condition \eqref{gauss} can be relaxed to \eqref{subexp}
if $\eta \ge N^{-1+\e}$ with any $\e>0$ at the expense of
a weaker bound on the right hand side of \eqref{P}, see Section 5 of \cite{ERSY2}.
The estimate also deterioriates if the energy is close to the edge,
see Proposition 4.1 of \cite{ERSY} for a more precise statement. 
Based upon our proofs, similar estimates were
given in \cite[Theorem 56]{TV} for energies in the bulk and
somewhat stronger bounds  in \cite[Theorem 1.7]{TV2} for the edge.

\bigskip

{\it Sketch of the proof.} For any $z=E+i\eta$, $\eta >0$, let
\be
    m(z) = m_N(z)= \frac{1}{N} \tr \frac{1}{H-z} = \frac{1}{N}\sum_{\al=1}^N
  \frac{1}{\lambda_\al -z}
\label{def:st}
\ee
be the Stieltjes transform of the empirical density of states
and let
$$
   m_{sc}(z) = \int \frac{\varrho_{sc}(x)}{x-z} \rd x
$$
be the Stieltjes transform of the semicircle law. Clearly $\varrho_\eta(E) = 
\frac{1}{\pi}\im \; m(z)$ gives the normalized density of states of $H$
around $E$ regularized on a scale $\eta$. Therefore it is sufficient to
establish the convergence of $m(z)$ to $m_{sc}(z)$ for small $\eta = \im \; z$.

The first step of the proof is to provide an
upper bound on $\cN_I$. Let $B^{(k)}$ denote the $(N-1)\times (N-1)$
minor of $H$ after removing the $k$-th row and $k$-th column. Let
 $\lambda_\al^{(k)}$, $\al=1,2, \ldots N-1$ denote the eigenvalues of $B^{(k)}$
and $\bu_\al^{(k)}$ denote its eigenvectors. Computing the $(k,k)$ 
diagonal element of the resolvent $(H-z)^{-1}$ we easily obtain the following expression
for $m(z)$
\be
   m(z) = \frac{1}{N}\sum_{k=1}^N \frac{1}{H-z}(k,k)
  = \frac{1}{N}\sum_{k=1}^N \Bigg[ h_{kk} -z-\frac{1}{N}
  \sum_{\al=1}^{N-1} \frac{\xi_\al^{(k)}}{\lambda_\al^{(k)}-z}\Bigg]^{-1},
\label{id}
\ee
where 
\be
\xi_\al^{(k)} = N |\ba^{(k)}\cdot \bu_\al^{(k)}|^2,
\label{xidef}
\ee
 and
$\ba^{(k)}$ is the $k$-th column of $H$ without the diagonal element $h_{kk}$. 
Taking the imaginary part, and using $\cN_I\le C\im \; m(z)$, we have
\be
  \cN_I \le CN\eta^2 \sum_{k=1}^N \Big| \sum_{\al\;: \; \lambda_\al^{(k)}\in I} 
  \xi_\al^{(k)}\Big|^{-1}.
\label{key}
\ee
It is an elementary fact that the eigenvalues of $H$ and $B^{(k)}$, for each fixed $k$,
are interlaced, i.e. the number of $\lambda_\al^{(k)}$ in $I$ is at least $\cN_I-1$.
For each fixed $k$ the random variables $\{ \xi_\al^{(k)} \; : \; \al=1, 2, \ldots N-1\}$
are almost independent and have expectation value one, thus
the probability of the event
$$
   \Omega_k : =   \Big\{ \sum_{\al\;: \; \lambda_\al^{(k)}\in I} 
  \xi_\al^{(k)} \le \delta (\cN_I-1)\Big\}
$$
is negligible for small $\delta$ \cite[Lemma 4.7]{ESY3}. On the complement
of all $\Omega_k$ we thus have from \eqref{key} that
$$
  \cN_I \le \frac{CN^2\eta^2}{\delta (\cN_I-1)},
$$
from which it follows that $\cN_I \le C N\eta$ with very high probability.

The second step of the proof is to establish that $m(z)$ and $m_{sc}(z)$ are
close. Let $m^{(k)}(z)$ denote the Stieltjes transform of the
empirical  distribution of the eigenvalues $\lambda_\al^{(k)}$ of $B^{(k)}$.
Then it follows from \eqref{id} that
\be
  m(z) = \frac{1}{N}\sum_{k=1}^N \frac{1}{h_{kk} - z- \big(
  1- \frac{1}{N}\big) m^{(k)}(z) - X_k}
\label{m}
\ee
holds, where 
$$
   X_k = \frac{1}{N} \sum_{\al=1}^{N-1} \frac{\xi_\al^{(k)}-1}{\lambda_\al^{(k)}-z}.
$$
Fixing the matrix $B^{(k)}$, we  view $X_k$ as a random variable of
the independent $\ba^{(k)}$ vector alone. Using again that the nominators $\xi_\al^{(k)}-1$
are almost independent and have zero expectation, we obtain that
$X_k$ is bounded by $(N\eta)^{-1}$ with high probability \cite[Lemma 6.1]{ESY3}.
The interlacing property guarantees that $m(z)$ and $m^{(k)}(z)$ are close.
Since  $h_{kk}$ is also small, we obtain from \eqref{m} that
\be
  m(z) = -\frac{1}{N}\sum_{k=1}^N \frac{1}{m(z)+z+\e_k}.
\label{mm}
\ee
where $\e_k$ are small with very high probability.
Note that the Stietljes transform of the semicircle law
is the solution of the equation
\be
   m_{sc}(z) = -\frac{1}{m_{sc}(z)+z}
\label{msc}\ee
that is stable away from the spectral edges, $z =\pm 2$.
Comparing the solution of \eqref{mm} and \eqref{msc} 
we obtain that $|m-m_{sc}|$ is small.
Strictly speaking, this argument applies only
for $\eta \ge (\log N)^4/N$ since  the smallness of
each $\e_k$ is guaranteed only apart from a set of
probability $e^{-c\sqrt{N\eta}}$ \cite[Lemma 4.2]{ESY3}
and there are $N$ possible values of $k$. On very short scale,
our proof uses an additional expansion
of  the denominators in \eqref{mm}
up to second order and we use that the expectation
of $X_k$, the main contribution to $\e_k$, vanishes
\cite[Section 6]{ESY3}.
 \qed

\bigskip

The second result concerns the delocalization of eigenvectors.
The motivation comes from the Anderson model.
In the infinite volume, the extended states regime is
usually characterized by the absolute continuity of the spectrum;
such characterization is meaningless for finite matrices.
However, the lack of concentration  of the eigenfunctions for
the  finite volume approximations of the Anderson
Hamiltonian is already a signature of the extended states regime.

If $\bv$ is an $\ell^2$-normalized eigenvector of $H$, then
the size of the $\ell^p$-norm of $\bv$, for $p>2$, gives
information about delocalization. 
Complete delocalization occurs
when $\|\bv \|_p \lesssim N^{-1/2+ 1/p}$ (note that  $\| \bv\|_p \ge 
CN^{-1/2+ 1/p}\|\bv\|_2$).
 The following result
shows that eigenvectors are fully delocalized with
a very high probability.

\begin{theorem}\cite[Corollary 3.2]{ESY3} Under the conditions
of Theorem \ref{thm:locsc}, for any $|E|<2$, fixed $K$ and $2<p<\infty$
we have
$$
   \P\Bigg\{ \exists \bv\; : \; H\bv=\lambda\bv, \; |\lambda-E|\le \frac{K}{N},
   \; \|\bv\|_2=1, \; \|\bv\|_p\ge MN^{-\frac{1}{2}+\frac{1}{p}}\Bigg\} 
  \le Ce^{-c\sqrt{M}}
$$
for $M$ and $N$ large enough.
\end{theorem}

The proof is an easy consequence of Theorem \ref{thm:locsc} and will be 
omitted here.

\bigskip

The local semicircle law asserts that the empirical density
on scales  $\eta\gg O(1/N)$  is close to the semicircle density.
On even smaller scales $\eta \le O(1/N)$, the emprical density fluctuates, but
its average, $\E \, \varrho_\eta (E)$, remains bounded
uniformly in $\eta$.  This is  
a type of Wegner estimate that plays a central role
in the localization theory of random Schr\"odinger operators.
In particular, it says that the probability of finding
at least one eigenvalue in an interval $I$ of size $\eta =\e/N$
is bounded by $C\e$ uniformly in $N$ and $\e\le 1$, i.e. no eigenvalue
can stick to any value.
Furthermore, if the eigenvalues were independent (Poisson process), then the probability
of finding $n=1, 2,3,\ldots $ eigenvalues in $I$ were proportional with
$\e^n$. For random matrices in the bulk of the
spectrum this probability is much smaller. 
This phenomenon is known
as level repulsion and the precise statement is the following:

\begin{theorem}\cite[Theorem 3.4 and 3.5]{ESY3}\label{thm:repul}
Suppose \eqref{gauss} holds and the measure $\nu$ is absolutely continuous
with a  strictly positive and smooth density. Let $|E| < 2$ and
$I= [E-\eta/2, E+\eta/2]$ with $\eta = \e/N$. Then  for any fixed $n$,
\be
   \P (\cN_I \ge n) \le
 \left\{  \begin{array}{cl} C_n\e^{n^2} &  \quad \mbox{[hermitian case]}\\
  C_n \e^{n(n+1)/2} & \quad\mbox{[symmetric case]}
 \end{array} \right.
\label{rep}
\ee
uniformly in $\e\le 1$ and for all sufficiently large $N$.
\end{theorem}

The exponents are optimal as one can easily see
from the Vandermonde determinant
in the joint probability density \eqref{expli} for unitary
ensembles. The sine kernel behavior \eqref{sine}
implies level repulsion (and even a lower bound on $\P (\cN_I \ge n)$),
 but usually not on arbitrarily small
scales since sine kernel is typically proven 
only as a weak limit (see \eqref{mainres} later).

\bigskip

{\it Sketch of the proof.} The starting point is formula
\eqref{id} together with 
$$
   \cN_I \le CN\eta \, \im\, m(E+i\eta).
$$
This implies
\be
 \cN_I
\leq C \eta\sum_{k=1}^N \frac{1}{(a_k^2 + b_k^2)^{1/2}}
\label{out}
\ee
with
$$
 a_k := \eta + \frac{1}{N}\sum_{\al=1}^{N-1}
\frac{\eta \xi_\al^{(k)}}{(\lambda_\al^{(k)}-E)^2 + \eta^2}, \qquad
b_k:= h_{kk} - E - \frac{1}{N}\sum_{\al=1}^{N-1}
\frac{ (\lambda_\al^{(k)}-E)
\xi_\al^{(k)}}{(\lambda_\al^{(k)}-E)^2 + \eta^2}\; ,
$$
where $a_k$ and $b_k$ are the imaginary and real
part, respectively, of the reciprocal of the summands in \eqref{id}
and $\xi_\al^{(k)}$ was defined in \eqref{xidef}.
The proof of  Theorem \ref{thm:locsc}  relied only on
the imaginary part, i.e.  $b_k$
in \eqref{out} was  neglected.
In the proof of Theorem \ref{thm:repul}, however, we
make an essential use of $b_k$ as well. Since typically
$1/N \lesssim |\lambda_\al^{(k)}-E|$, we
note that $a_k^2$ is much smaller
than $b_k^2$ if $\eta\ll 1/N$
and this is the relevant regime for the Wegner estimate and
for the level repulsion.

Assuming a certain smoothness condition on the distribution $\rd \nu$,
 the distribution of
the variables $\xi_\al^{(k)}$ will also be smooth even 
if we fix an index $k$ and we condition on the minor $B^{(k)}$, i.e. if we fix
the eigenvalues $\lambda_\al^{(k)}$ and the eigenvectors $\bu_\al^{(k)}$.
 Although the random variables $\xi_\al^{(k)}= N|\ba^{(k)}\cdot \bu_\al^{(k)}|^2$
are not independent for different $\al$'s, they are sufficiently
decorrelated so that the distribution of $b_k$ inherits some 
smoothness from $\ba^{(k)}$. Sufficient smoothness on the distribution of $b_k$
makes the expectation value $(a_k^2 + b_k^2)^{-p/2}$
finite for any $p>0$. This will give a bound on the $p$-th moment
on $\cN_I$ which will imply \eqref{rep}. 

We present this idea for hermitian matrices and for the simplest
case $k=1$. {F}rom \eqref{out} we have
$$
   \P (\cN_I\ge 1) \le \E \, \cN_I^2 \le C(N\eta)^2 \E \frac{1}{a^2_1 + b^2_1} .
$$
Dropping the superscript $k=1$ and introducing the notation
$$
    d_\al = \frac{N(\lambda_\al -E)}{N^2 (\lambda_\al-E)^2 +\e^2},
  \qquad c_\al =\frac{\e}{N^2 (\lambda_\al-E)^2 +\e^2},
$$
we have
\be
   \P (\cN_I\ge 1) \le C\e^2 \, \E \Bigg[ \Big( \sum_{\al=1}^{N-1}
   c_\al\xi_\al\Big)^2 + \Big( h - E - \sum_{\al=1}^{N-1}
   d_\al\xi_\al\Big)^2 \Bigg]^{-1}.
\label{PN}
\ee
From the local semicircle law we know that with very high probability,
there are several eigenvalues 
$\lambda_\al$ within a distance of $O(1/N)$ of $E$. Choosing four such
eigenvalues, we can guarantee that  for some index $\gamma$ 
\be
  c_\gamma, c_{\gamma+1} \ge C\e, \quad  d_{\gamma+2}, d_{\gamma+3} \ge C
\label{cbound}\ee
for some positive constant $C$. If $\xi_\al$'s were indeed independent 
and distributed according to the square of a complex random variable $z_\al$
with a smooth and decaying density $\rd\mu(z)$ on the complex plane, then
the expectation in \eqref{PN} would be bounded by
\be
   \sup_{E}\int \frac{1}{ \big(c_\gamma |z_\gamma|^2 + c_{\gamma+1} |z_{\gamma+1}|^2\big)^2
  +  \big( E - d_{\gamma+2} |z_{\gamma+2}|^2 -d_{\gamma+3} |z_{\gamma+3}|^2 \big)^2  }
  \prod_{j=0}^3 \rd \mu(z_{\gamma+j}).
\label{exint}\ee
Simple calculation shows that this integral is bounded by $C\e^{-1}$ 
assuming the lower bounds \eqref{cbound}. 
Combining this bound with \eqref{PN}, we obtain \eqref{rep} 
for $n=1$. The proof for the general $n$ goes by induction.
The difference between the hermitian and the symmetric cases
manifests itself in the fact that $\xi_\al$'s are squares
of complex or real variables, respectively. This gives different estimates
for integrals of the type \eqref{exint}, resulting in different
exponents in \eqref{rep}. \qed

\section{Sine kernel universality}

Let $f(\lambda_1, \lambda_2, \ldots ,\lambda_N)$ denote
the symmetric joint density function of the eigenvalues of the $N\times N$
Wigner matrix $H$. For any $k\ge 1$ we define the $k$-point
correlation functions (marginals) by
$$
   p^{(k)}_N(\lambda_1, \ldots ,\lambda_k) = \int_{\R^{N-k}}
  f(\lambda_1, \lambda_2, \ldots , \lambda_N)\rd\lambda_{k+1}\ldots 
 \rd \lambda_N.
$$
We will use the notation $p^{(k)}_{N, GUE}$ and  $p^{(k)}_{N, GOE}$ 
for the correlation functions of the GUE and GOE ensembles.

We consider the rescaled correlation functions about a
fixed energy $E$ under a scaling that guarantees that the
local density is one. The sine-kernel universality for the GUE
ensemble states that
the rescaled correlation functions converge weakly
to the determinant of the sine-kernel, $K(x)= \frac{\sin \pi x}{\pi x}$, i.e.
\be
   \frac{1}{[\varrho_{sc}(E)]^k} p^{(k)}_{N, GUE}
  \Big( E + \frac{x_1}{N\varrho_{sc}(E)},\ldots
 E + \frac{x_k}{N\varrho_{sc}(E)}\Big) \to 
  \det \big( K(x_\ell - x_j)\big)_{\ell,j=1}^k
\label{GUEuniv}
\ee
as $N\to\infty$
for any fixed energy $|E|<2$ in the bulk of the spectrum
\cite{MG, D}.
Similar result holds for the GOE case; the sine kernel
being replaced with a similar but somewhat more complicated
universal function, see \cite{M}. Our main result is that
universality \eqref{GUEuniv} holds for general hermitian or symmetric
Wigner matrices after averaging in the energy $E$:

\begin{theorem}\label{mainthm} \cite{ESY4} Let $H$ be an $N\times N$ symmetric or hermitian
Wigner matrix with normalization  defined at the beginning of 
Section \ref{sec:sc}. Suppose that the distribution $\nu$ of the matrix elements
has subexponential decay \eqref{subexp}. Let $k\ge 1$ and 
  $O: \R^k \to \R$ be a continuous, compactly supported function. 
Then for any $ |E| < 2 $, we have 
\begin{equation}\label{mainres}
\begin{split}
\lim_{\delta\to 0} \lim_{N \to \infty}  & \frac{1}{2\delta} \int_{E-\delta}^{E+\delta}  \rd v
 \int_{\R^k}  \rd\alpha_1 
\ldots \rd\alpha_k \; O(\alpha_1,\ldots,\alpha_k) 
\\ &\times  \frac{1}{[\varrho_{sc}(v)]^k} \Big ( p_N^{(k)}  - p_{N, \#} ^{(k)} \Big )
  \Big (v+\frac{\alpha_1}{N\varrho_{sc}(v)}, 
\ldots, v+\frac{\alpha_k}{N \varrho_{sc}(v)}\Big) =0,
\end{split} 
\end{equation}
where $\#$ stands for GOE or GUE for the symmetric or hermitian cases, respectively.
\end{theorem}

For the {\it hermitian case}, the first result on universality beyond the GUE
was due to Johansson \cite{J} (based upon \cite{BH}) under the condition that $\nu$ has
a Gaussian component with a positive variance independent of $N$.
 His method was extended in \cite{BP} to  Wishart matrices. 
The variance of the necessary Gaussian component was reduced to 
$N^{-3/4+\e}$ in \cite{ERSY} under the additional
technical assumptions that the measure $\nu$ is smooth and it satisfies the
logarithmic Sobolev inequality. The local statistics was identified
via orthogonal polynomials.
The Gaussian component  assumption was
first removed completely in \cite{ERSY2} under the condition that the
density of the probability measure $\nu$ is positive and it possesses
a certain number of derivatives. Shortly after \cite{ERSY2}
appeared on the arXiv, the same result using a different method
has been posted \cite{TV}
without any regularity condition on $\nu$ provided
that the third moment vanishes and $\nu$ is supported on
at least three points. Combining the two methods,
all conditions on $\nu$ apart from the subexponential decay \eqref{subexp}
were removed in a short joint paper \cite{ERSTVY}.

The methods of \cite{ERSY2} and \cite{TV} both 
 rely on the explicit formula of Br\'ezin and Hikami \cite{BH}, exploited
also in \cite{J}, for the correlation functions of the Wigner matrix with Gaussian
convolution.  This formula reduces the problem to a saddle point analysis.
The saddle points are identified by solving an equation involving
the Stieltjes transform $m_N(z)$ \eqref{def:st} with $\eta = \im \, z$ corresponding
to the variance of the Gaussian component: precise information on $m_N(z)$
for a smaller $\eta$ implies that a smaller Gaussian component is sufficient.

In our work \cite{ERSY2} we used the 
convergence of $m_N(z)$ to $m_{sc}(z)$ for very small $\eta=  N^{-1+\e}$ 
established along the proof of Theorem \ref{thm:locsc}. To remove 
this tiny Gaussian component, we have compared the local
eigenvalue statistics of a given Wigner matrix $H$ with that
of $\wh H_s + sV$ for which the saddle point analysis applies.
Here $s^2= \eta = N^{-1+\e}$ and the new Wigner matrix $\wh H_s$  was chosen such that the 
law of $\wh H_s + sV$ be very close to  $H$. Since Gaussian
convolution corresponds to running a heat flow on the matrix
elements, $\wh H_s$ could, in principle, be obtained by
running the {\it reverse heat flow} on the elements of $H$.
Although the reverse heat flow is undefined for most initial
conditions, one can
construct an appoximation  to the reverse heat flow that
is well defined and yields $\wh H_s$ with a required
precision  assuming sufficient smoothness on $\nu$.
Technically, we use Ornstein-Uhlenbeck process
instead of the heat flow to keep the variance constant.
We also mention that the result of \cite{ERSY2} is valid
for any fixed energy $E$, i.e. $\rd v$ averaging
in \eqref{mainres} is not necessary.

Tao and Vu \cite{TV} have directly compared local statistics
of the Wigner matrix $H$ and that of the matrix with order one
Gaussian component for which Johansson has already proved universality.
Their main technical result \cite[Theorem 15]{TV} states
that the local eigenvalue statistics
of two Wigner matrices coincide as long as the first four
moments of their single site
distributions match. It is then an elementary 
lemma from probability theory (\cite[Corollary 23]{TV} based upon
\cite{CF}) to match to order four a given random variable 
with another random variable with a Gaussian component.

The proof of Theorem \ref{mainthm} for the {\it symmetric} case
requires a new idea since the formula of Br\'ezin and Hikami is
not available. While the four moment theorem of \cite{TV} also
applies to this case, there is no reference ensemble available. 
In the next sections we describe our new approach that
proves universality for both hermitian and symmetric 
matrices without relying on any explicit formulae.

\section{Dyson Brownian motion}

The joint distribution of the 
eigenvalues $\bx= (x_1, x_2, \ldots , x_N)$ 
of the Gaussian ensembles 
is given by the following measure
\be\label{H}
\mu=\mu_N(\rd{\bf x})= 
\frac{e^{-\cH({\bf x})}}{Z_\beta}\rd{\bf x},\qquad \cH({\bf x}) =  
 N \left [ \beta \sum_{i=1}^N \frac{x_{i}^{2}}{4} -  \frac{\beta}{N} \sum_{i< j} 
\log |x_{j} - x_{i}| \right ]
\ee
where 
 $\beta=1$ for GOE and $\beta=2$ for GUE. 
For definiteness, we consider the $\beta=1$ GOE case
and we assume that the eigenvalues are ordered, i.e.
$\mu$ is restricted to 
$\Sigma_N = \{ \bx\in \R^N\; : \; x_1 < x_2<  \ldots < x_N\}$.

Suppose the matrix elements evolve according to 
 the  Ornstein-Uhlenbeck process  on $\bR$, i.e. 
the density of their distribution $\nu_t = u_t(x) \rd x$ satisfies
\be\label{ou}
\partial_{t} u_t =  \cL u_t, \quad 
   \cL  = \frac{1}{2}\frac{\pt^2}{\pt x^2}
- \frac{ x}{2} \frac{\pt}{\pt x}.
\ee
The Ornstein-Uhlenbeck  process \eqref{ou} induces a
stochastic process, the Dyson Brownian motion, 
 on the eigenvalues with a  generator given by 
\be
 L=   \sum_{i=1}^N \frac{1}{2N}\partial_{i}^{2}  +\sum_{i=1}^N
 \Bigg(- \frac{\beta}{4} x_{i} +  \frac{\beta}{2N}\sum_{j\ne i} 
\frac{1}{x_i - x_j}\Bigg) \partial_{i}
\label{L}
\ee
acting on $L^2(\mu)$. The measure $\mu$ is invariant
and reversible with respect to the dynamics generated by $L$.
Let
\be
D(f) = -\int  f L f  \rd \mu =  \sum_{j=1}^N \frac{1}{2N}
\int (\partial_j f)^2 \rd \mu
\label{def:dir}
\ee
be the corresponding Dirichlet form. 
Denote the distribution of the eigenvalues  at  time $t$
 by $f_t ({\bf x})\mu(\rd {\bf x})$.
 Then $f_t$ satisfies 
\be\label{dy}
\partial_{t} f_t =  L f_t
\ee
with initial condition $f_0$ given by the eigenvalue density of the
Wigner ensemble. Dyson Brownian motion is the
 corresponding system of stochastic differential equations
 for the eigenvalues ${\bf x}(t)$ that is  given by  (see, e.g.
Section 12.1 of \cite{G})
\be\label{sde}
  \rd  x_i  =    \frac{\rd B_i}{\sqrt{N}} +  \left [ -  \frac{\beta}{4}
 x_i+  \frac{\beta}{2 N}\sum_{j\ne i} 
\frac{1}{x_i - x_j}  \right ]  \rd t, \qquad 1\leq i\leq N,
\ee
where $\{ B_i\; : \; 1\leq i\leq N\}$ is a collection of
independent Brownian motions.  Note that the equations \eqref{dy}
and \eqref{sde} are defined for any $\beta\ge 1$, independently
of the original matrix models. 
Our main technical result (Theorem \ref{thmM}) holds for
general $\beta\ge 1$.

\section{Local Relaxation Flow}

The Hamiltonian of the
 invariant measure $\mu$ of the Dyson Brownian motion
is convex, with Hessian  bounded from below 
$$
  \mbox{Hess} \, \cH \ge \frac{\beta N}{2}
$$
on the set $\Sigma_N$. By the Bakry-Emery criterion,
this guarantees that $\mu$ satisfies the logarithmic
Sobolev inequality and the relaxation time to 
equilibrium is of order one (note the additional
$1/N$ factor in the Dirichlet form \eqref{def:dir} that rescales time).

We now introduce the local relaxation measure, which has the
local statistics of GOE (or GUE)
but generates  a faster decaying dynamics.
Let $\gamma_j$ be the semicircle location of
the $j$-th eigenvalue, i.e.
$$
   \gamma_j= n_{sc}^{-1} (j/N), \qquad 
  n_{sc}(E) := \int_{-\infty}^E \varrho_{sc}(x) \rd x.
$$
We fix a regularization parameter $\eta \ll 1$ and
we replace the interaction potential between $x_j$ and
far away particles by a regularized mean field potential
\be\label{Wj}
W_j(x) =  -\frac \beta N  \sum_{k: |k-j| \ge N \eta} \log (|x - \gamma_k|+\eta) 
%\quad \text { if }\quad   x\in I_j :=   (\gamma_j^-, \gamma_j^+) .
\ee
Strictly speaking, $W_j(x)$ is defined by this formula
only in an interval of size $N\eta$ about $\gamma_j$ 
and we use a quadratic extension beyond, but we leave this technicality
aside.

The local relaxation measure $\om_N=\om $ is a Gibbs measure 
defined by the Hamiltonian
\[
\wt\cH = N   \sum_{j=1}^N \left \{  \beta  \frac {x_j^2} 4  +
  W_j (x_j) \right \}  +  \beta \sum_{i<j } 
  \log |x_i- x_j| - \frac \beta 2  
\sum_i \sum_{j:  |j-i| > N \eta}   \log( |x_i- x_j|+\eta) .
\] 
We often write $\om =\psi\mu$ where $\psi$ is the Radon-Nykodim derivative.
The local relaxation flow is defined to be  the reversible dynamics w.r.t. 
$\omega$ characterized by the generator  $\tilde L$ defined by 
\be\label{Lt}
 \int f \tilde L  g \rd \omega = - \frac 1 {2N} \sum_j  
\int \partial_j f \partial_j g \rd \omega .
\ee
Explicitly,  $\wt L$ is given by  
\be\label{tl}
L = \wt L + \sum_j b_j \partial_j, \quad
 b_j =   \frac 1 {N} \sum_{k: |k-j|> N \eta}
  \frac {\mbox{sgn}(x_j-x_k)}{|x_j-x_k|+\eta}
 + W'_j(x_j),
\ee 
Simple calculation shows that the mean field potential is 
uniformly convex with
\be\label{5.9}
\inf_j \inf_{ x \in \bR}     W_j^{\prime \prime}(x) 
  \ge c \eta^{-1/3}.
\ee
This will guarantee that the relaxation time to equilibrium $\om$
for the $\wt L$ dynamics is of order $\eta^{-1/3}$.

We recall the definition of the relative entropy of
 with respect to any probability measure $\rd\lambda$
\[
S_\lambda(f) = \int f\log f \rd \lambda, \qquad
S_\lambda (f|\psi)=  \int f \log (f/\psi) \rd \lambda
\]

Our main technical result is the following theorem that states that
the relaxation time $\tau$ for specific local observables is much shorter
than order one.

\begin{theorem}[Universality of Dyson Brownian Motion for Short  Time]\label{thmM} 
Suppose that $S_\mu(f_0|\psi) \le C N^m$ for some $m$
 fixed. Let
$ \tau = \eta^{1/3} N^\e $ with some $\e>0$
and assume that $\eta \ge N^{-3/55+\e}.$
Assume that  there is a positive number $\Lambda$ such that 
\be\label{bbound} 
\sup_{0\le t \le \tau} N \sum_j \int b_j^2 f_t \rd \mu  \le C   \eta^{-2} 
  \Lambda  .  
\ee
Let  $G$ be a bounded smooth function with compact support.
Then for any  fixed $n\ge 1$ and $J \subset [1, \ldots, N]$ we have 
\[
\Big | \int \frac 1 N \sum_{i\in J}  G(N(x_i - x_{i +n})) f_\tau  \rd \mu - 
\int \frac 1 N \sum_{i\in J} G(N(x_i - x_{i +n})) \rd \mu \Big | \le 
\sqrt{ \frac{C\Lambda}{ N^{1-\e} \eta^{5/3}}}.
\]
\end{theorem}

\bigskip We emphasize that Theorem \ref{thmM} applies to {\it all $\beta\ge1$ ensembles} 
and the only assumption 
concerning  the distribution $f_t$
is in  (\ref{bbound}).  In case of the original Wigner ensembles $\beta=1, 2$,
the critical  constant $\Lambda$  can be estimated under an  additional 
assumption.

\begin{lemma}\label{lm:bbound} Let $f_0$ be the joint density of the
eigenvalues of a Wigner matrix.
Suppose  that the measure $\rd\nu$ of its single site distribution
 satisfies the logarithmic Sobolev
inequality.  Then the constant $\Lambda$ in 
\eqref{bbound} can be estimated as
\be\label{bbound1} 
  \Lambda    \le C_{\sigma} \eta^{-2} N^{4/5+\sigma} 
\ee
for any $\sigma>0$.
\end{lemma}

For the proof of this lemma, we can estimate $b_j$ as
 \be\label{bj}
 |b_j| \le    \; \frac 1 N \sum_{k \; : \; |k-j|> N \eta} 
\left | \frac {\mbox{sgn}(x_j-x_k)} {|x_j-x_k|+\eta}
 - \frac {\mbox{sgn}(x_j -\gamma_k) }{ |x_j -\gamma_k|+\eta} \right |
\le C\eta^{-2} \frac{1}{N}\sum_{k=1}^N |x_k-\gamma_k|
\ee
as long as $x_k$ is sufficiently near $\gamma_k$ so that
$\mbox{sgn} (x_j -\gamma_k) =\mbox{sgn}(\gamma_j-\gamma_k)$ holds for $|j-k|>N\eta$.
The average difference between $x_k$ and $\E x_k$ can be estimated
using the logarithmic Sobolev inequality for $\nu$. The average
of $|\E x_k -\gamma_k|$ is estimated in Proposition 4.2 of \cite{ESY3}
that was a consequence of the local semicircle law.
Combining these results with information on the lowest
and largest eigenvalues \cite{Vu}, we can show that 
$\frac{1}{N} \sum_k |x_k-\gamma_k|\le N^{-3/5+\e}$ and this yields
\eqref{bbound1}. \qed

\bigskip

Combining Lemma \ref{lm:bbound} with Theorem  \ref{thmM} and choosing
$\eta$ appropriately, we see that the local eigenvalue
statistics of $f_\tau$ with $\tau \ge N^{-1/55 +\e}$
coincides with that of the global equilibrium measure, i.e.
with GOE or GUE. For hermitian matrices, 
the same statement was already proven in \cite{ERSY}
even for $\tau \ge N^{-1+\e}$ by using Br\'ezin-Hikami formula, but
the current approach is purely analytical and it applies
to symmetric matrices as well.  Using the reverse heat flow argument,
we can show that the local statistics of $f_0$ is also given
by GOE or GUE assuming that the initial distribution $\nu$
is sufficiently smooth. The smoothness condition and 
the additional requirement  that $\nu$ satisfies the
logarithmic Sobolev inequality can be removed by applying
the four moment theorem of \cite{TV}.

\section{Proof of Theorem \ref{thmM}}

We first list the key new ideas of 
behind the proof of Theorem \ref{thmM},
then we formulate the corresponding results.

\begin{enumerate}

\item[I.] The key concept is the introduction of 
the local relaxation flow \eqref{Lt} which has the following two properties: 
(1)  The invariant measure for this flow, the local relaxation measure $\omega$
has  
the same local eigenvalue statistics as the  GOE or GUE. 
(2)   The relaxation time of the local relaxation flow is much shorter 
than that of the DBM, which is of order one.

\item[II.]  Suppose we have a density
 $q$ w.r.t. $\omega$ that evolves with the local relaxation 
flow. Then, by differentiating  the Dirichlet form 
w.r.t. $\omega$  we will prove that
 the difference between the local  statistics 
of $q\omega$ and $\omega$ can be estimated in terms of the Dirichlet
 form of $q$ w.r.t. $\omega$. 
Hence if the Dirichlet form is small, the local statistics of
 $q \omega$ is independent of $q$.

\item [III.] It remains to show that  the Dirichlet form of 
$q=f_t \mu$ w.r.t. $\omega$ is small for $t$ sufficiently large 
(but still much less than order one). To do that,
 we study the evolution of the entropy of 
$f_t \mu$ relative to $\omega$. This provides estimates on
 the entropy and Dirichlet form which   serve as
inputs for the Step II to  conclude the universality.

\end{enumerate}

The first ingredient to prove Theorem \ref{thmM} is the analysis of 
the local relaxation flow which  satisfies the logarithmic
Sobolev inequality and the following 
dissipation estimate.

\begin{theorem}\label{thm2} 
 Suppose \eqref{5.9} holds. 
Consider the equation 
\be
\partial_t q_t=\tilde L q_t
\label{dytilde}
\ee
with reversible measure $\omega$.
Then we have the following estimates
\be\label{0.1}
\partial_t D_{\omega}( \sqrt {q_t}) \le - C \eta^{-1/3} D_{\omega}( \sqrt {q_t}) -  
 \frac{1}{2N^2}  \int    \sum_{ |i-j| \le N \eta} 
 \frac 1 {(x_i - x_j)^2}     ( \pt_i \sqrt{ q_t} - \pt_j\sqrt {q_t} )^2 \rd \omega , 
\ee
\be\label{0.2}
 \frac{1}{2N^2} \int_0^\infty  \rd s  \int    \sum_{ |i-j| \le N \eta} 
 \frac 1 {(x_i - x_j)^2}     (\pt_i\sqrt {q_s} - \pt_j\sqrt {q_s} )^2 \rd \omega
 \le D_{\omega}( \sqrt {q_0}) 
\ee
and the logarithmic Sobolev inequality 
\be\label{lsi}
  S_{\omega}(q)\le C \eta^{1/3}  D_{\omega}( \sqrt {q})
\ee
with a universal constant $C$.
Thus the time to equilibrium is of order $\eta^{1/3}$: 
\be\label{Sdecay}
   S_{\omega}(q_t)\le e^{-Ct \eta^{-1/3}} S_\omega(q_0).
\ee
\end{theorem}

 The proof  follows the standard argument in \cite{BE} (used
in this context in \cite{ERSY}). The key input is 
the following lower bound on the Hessian of $\wt \cH$
\be\label{convex}
\frac{1}{2N^2}\;
 \Big\langle \bv , (\nabla^2 \wt\cH)\bv\Big\rangle 
\ge   C \eta^{-1/3}    \frac 1 N \|\bv\|^2 + 
 \frac{1}{2N^2}  \sum_{ |i-j| \le N \eta}  \frac 1 {(x_i - x_j)^2} (v_i - 
v_j)^2  .
\ee
The first term is due to convexity of the mean field potential
\eqref{5.9}. The second term comes from the additional convexity of
the local interaction and it corresponds to  ``local  
Dirichlet form dissipation". 
The estimate \eqref{0.2} on this additional term 
plays a key role in the next theorem.

\begin{theorem}\label{thm3}
Suppose that the density $q_0$ satisfies $S_\omega(q_0)\le CN^m$ with some $m>0$ fixed.
Let  $G$ be a bounded smooth function with compact support
and let $J\subset \{ 1, 2, \ldots , N\}$. Set $\tau = \eta^{1/3} N^\e$.
Then  for any  $n\ge 1$  fixed we have 
\be\label{diff}
\begin{split}
\Big| \int \frac 1 N \sum_{i\in J} G(N(x_i - x_{i +n})) \rd \omega - 
\int \frac 1 N \sum_{i\in J}&  G(N(x_i - x_{i +n}))  q_0 \rd \omega \Big| \\
 & \le C\sqrt { \frac { D_\omega(\sqrt {q_0}) \tau  } {N} } +  Ce^{-cN^{\e}} .
\end{split}
\ee
\end{theorem} 

{\it Sketch of the proof.} Let $q_t$ satisfy
\[
\partial_t q_t = \tilde L q_t
\]
with an initial condition $q_0$. 
Thanks to the exponential decay of the entropy on time
scale $\tau\gg \eta^{1/3}$, see \eqref{Sdecay},  difference between 
the local statistics  w.r.t $q_\tau \om $ and $q_\infty \om =\om$ is subexponentially 
small in $N$.
To compare $q_0$ with $q_\tau$, 
by differentiation, we have
\[
\int \frac 1 N \sum_i G(N(x_i - x_{i +n}))q_\tau \rd \omega -
 \int \frac 1 N \sum_i G(N(x_i - x_{i +n}))q_0 \rd \omega \qquad\qquad\qquad\qquad
\]
\[
\qquad\qquad\qquad = \int_0^\tau  \rd s \int  \frac 1 N \sum_i G'(N(x_i - x_{i +n}))
  [\pt_i q_s - \pt_{i+n}q_s]  \rd \omega.
\]
{F}rom the Schwarz inequality and $\pt q = 2 \sqrt{q}\pt\sqrt{q}$
 the last term is bounded by 
\begin{align}\label{4.1}
& 2\left [ \int_0^\tau  \rd s \int  \frac 1 {N^2 } \sum_i 
  \frac 1 {(x_i - x_{i+n})^2}  [ \pt_i\sqrt {q_s} - 
\pt_{i+n}\sqrt {q_s}]^2  \rd \omega \right ]^{1/2} \nonumber \\
& \times  \left [   \int_0^\tau  \rd s \int  
 \sum_i G'(N(x_i - x_{i +n})) ^2  (x_i - x_{i+n})^{2}  q_s \rd \omega
\right ]^{1/2} 
\le  C\sqrt { \frac { D_\omega(\sqrt {q_0}) \tau } N },
\end{align}
where we have used \eqref{0.2} and 
 $G'(N(x_i - x_{i +n})) ^2  (x_i - x_{i+n})^{2} \le  C /N^2$. \qed

\bigskip 
Notice if we use only the entropy dissipation and Dirichlet form, 
the main term on the right hand side of \eqref{diff} will become $\sqrt {S \tau}$. 
Hence by exploiting the local Dirichlet form dissipation
coming from the second term on the r.h.s. of \eqref{0.1},
 we gain the crucial  factor $N^{-1/2}$
in the estimate.

\bigskip

The final  ingredient to prove Theorem \ref{thmM} is the following entropy and 
Dirichlet form estimates.
 
\begin{theorem}\label{thm1}
  Suppose the assumptions 
of Theorem \ref{thmM} hold. Let $\tau = \eta^{1/3} N^\e$ and 
let $g_t = f_t/\psi$ so that 
$S_{\mu} (f_t|\psi) =  S_{\omega} (g_t)$.
 Then the entropy and the Dirichlet form satisfy the estimates: 
\be\label{1.3}
  S_{\omega} (g_{\tau/2}) \le  
   C  \eta^{-5/3}   \Lambda, \qquad
 D_\omega (\sqrt{g_\tau})
 \le C\eta^{-2}\Lambda.
\ee
\end{theorem}

{\it Sketch of the proof.}  Recall that  $\pt_t f_t = Lf_t$.
The standard estimate on the entropy of $f_t$ with respect
to the invariant measure  is obtained by differentiating
it twice and using the logarithmic Sobolev inequality. The entropy 
and the Dirichlet form in \eqref{1.3} are, however, computed 
with respect to the measure $\om$. This yields the additional
second  term in the following identity  \cite{Y} that holds for
 any probability density
$\psi_t$:
$$
\partial_t  S_\mu(f_t|\psi_t) = - \frac 2 {  N } \sum_{j} \int (\partial_j
\sqrt {g_t})^2  \, \psi_t \, \rd\mu
+\int g_t(L-\partial_t)\psi_t \, \rd\mu \ ,
$$
 where $g_t = f_t/\psi_t$. In our application we set $\psi_t = \psi =\om/\mu$,
hence we have
$$
\pt_t S_\omega(g_t)
= - \frac 2 {  N } \sum_{j} \int (\partial_j
\sqrt {g_t})^2  \, \rd\omega
+\int   \wt L g_t  \, \rd\omega+  \sum_{j} \int  b_j \partial_j
g_t   \, \rd\omega.
$$
Since $\om$ is invariant, the middle term on the right hand side vanishes, and
  from the Schwarz inequality 
\be\label{1.1}
\pt_t S_\omega(g_t) \le  -D_{\omega} (\sqrt {g_t})
+   CN  \sum_{j} \int  b_j^2  g_t   \, \rd\omega.
\ee
Together with  \eqref{lsi}  and \eqref{bbound},  we have 
\be\label{1.2new}
\partial_t  S_\omega(g_t) \le  - C\eta^{-1/3}  S_\omega(g_t) + 
   C   \eta^{-2}  \Lambda.
\ee
which, after integrating it from $t=0$ to $\tau/2$,
 proves the first inequality in  \eqref{1.3}.
The second inequality can be obtained
from integrating  \eqref{1.1}
from $t=\tau/2$ to $t=\tau$ and using
 the monotonicity of the Dirichlet form in time. \qed

\bigskip

Finally, we sketch the proof of Theorem \ref{thmM}.
With the choice of
$\tau= \eta^{1/3} N^\e$
and $q_0=  f_\tau/\psi$, 
Theorems \ref{thm2}, \ref{thm3} and \ref{thm1} directly imply
\be\label{diff1}
\begin{split}
\Big| \int \frac 1 N \sum_{i\in J} G(N(x_i - x_{i +n})) f_\tau\rd \mu - 
\int \frac 1 N \sum_{i\in J}&  G(N(x_i - x_{i +n}))   \rd \omega \Big| \\
 & \le \sqrt{ \frac{C\Lambda}{ N^{1-\e} \eta^{5/3}}}
 +  Ce^{-cN^{\e}} ,
\end{split}
\ee
i.e. the local statistics of $f_\tau \mu$ and $\om$ are close.
Clearly, equation  \eqref{diff1}  also holds for the special choice
  $f_0= 1$ (for which $f_\tau=1$), i.e. local statistics of $\mu$ and
$\om$ can also be compared. This completes the proof of Theorem \ref{thmM}.
\qed

\end{document}